\documentclass[twocolumn,showpacs,preprintnumbers,amsmath,amssymb]{revtex4}
\usepackage{graphicx}
\usepackage{color}
\begin{document}

\title{A Naturally Minute Quantum Correction to the Cosmological Constant Descended from the Hierarchy}

\author{Shu-Heng Shao$^{1,3}$}\email{b95202055@ntu.edu.tw}

\author{Pisin Chen$^{1,2,3,4}$}\email{pisinchen@phys.ntu.edu.tw}
\affiliation{1. Department of Physics, 
National Taiwan University, Taipei 10617, Taiwan, R.O.C.\\
2. Graduate Institute of Astrophysics, 
National Taiwan University, Taipei 10617, Taiwan, R.O.C.\\
3. Leung Center for Cosmology and Particle Astrophysics, National Taiwan University, Taipei 10617, Taiwan, R.O.C.\\
4. Kavli Institute for Particle Astrophysics and Cosmology,
SLAC National Accelerator Laboratory, Menlo Park, CA 94025, U.S.A.}

\date{\today}
\begin{abstract}
We demonstrate that a naturally small quantum correction, or the Casimir energy, to the cosmological constant can arise from a massive bulk fermion field in the Randall-Sundrum model. Under the assumption that the ground value of four-dimensional effective cosmological constant is zero, which can only be attained by some other means, we show that its quantum correction originated from the Casimir energy, that is the vacuum energy associated with the non-trivial topology of the extra dimension, can be as small as the observed dark energy scale without fine-tuning of the bulk fermion mass. To ensure the stabilization of the system, we discuss two stabilization mechanisms under this setup. It is found that the Goldberger-Wise mechanism can be successfully introduced in the presence of a massive bulk fermion, without spoiling the smallness of the quantum correction.\end{abstract}

\maketitle

\section{Introduction}
The hierarchy between the Planck scale $M_{Pl}\sim10^{19}$ GeV and the electroweak scale $M_{EW}\sim$ TeV has long been a puzzle in high energy physics. In the past decade, there has been two famous solutions to the hierarchy problem, both involving the idea of extra dimensions. The first one is the Arkani-Hamed-Dimopoulos-Dvali (ADD) model \cite{ADD}, and the second is the Randall-Sundrum (RS) model \cite{RS}. In this Letter we shall only focus on the latter. In RS model, two parallel flat 3-branes, with equal but opposite brane tensions, are embedded in a five-dimensional Anti-de-Sitter (AdS) bulk. The RS metric reads
\begin{equation}
ds^2=e^{-2k r_c|\phi|}\eta_{\mu\nu}dx^\mu dx^\nu+r_c^2d\phi^2,\label{metric}
\end{equation}
where $k$ is the AdS curvature of the order $M$, the five-dimensional Planck scale, and $r_c$ is the radius of the $S^1/Z_2$ orbifold. $x^\mu$ are the four-dimensional coordinates and $\phi$ is the extra dimension coordinate ranging from $-\pi$ to $\pi$. The hidden brane is located at $\phi=0$ while the visible brane at $\phi=\pi$, on which the standard model fields reside. Under this construction, the hierarchy problem is naturally solved via the warp factor $e^{-\pi k r_c}$ along the fifth dimension without fine-tuning of the parameters. Specifically, the electroweak scale on the visible brane can be naturally bridged with the Planck scale on the hidden brane if $kr_c\simeq 12$.

There is another mysterious hierarchy problem in physics. The accelerating expansion of the present universe \cite{SN} implies that the cosmological constant (CC) may be nonzero but minute, which corresponds to an energy density $\rho_{obs}\equiv M_{CC}^4\sim (10^{-3}\mbox{eV})^4$. Since the quantum vacuum energy naturally contributes to CC, which is at the Planck scale, the long standing cosmological constant problem (CCP) \cite{CC} can be casted as a naturalness problem: Why is $\rho_{obs}$ smaller than the Planck energy density $\rho_{Pl}\sim (10^{19}\mbox{GeV})^4$ by $\sim124$ orders of magnitude?  

The CCP is actually more severe than the apparent minute value of CC. In the context of the quantum field theory, each field contributes a vacuum energy at the cut-off scale, say, the Planck scale. This vacuum energy could be dropped without harm for the quantum field theory in flat spacetime, since the interaction between particles is governed by the gradient of the potential, rather than its absolute value. On the contrary, in general relativity \textsl{all} forms of energy contribute to the dynamics of the spacetime, hence the vacuum energy gravitates in the same way as the cosmological constant. In addition, the spontaneous symmetry breaking of the quantum fields also gives rise to vacuum energy. Adding all these huge contributions from the quantum field theory, together with the bare cosmological constant, one is supposed to reduce the sum to the observed cosmological constant scale $\rho_{obs}\sim (10^{-3}\mbox{eV})^4$. It is logically possible that the bare cosmological constant manages itself in an extremely delicate manner so that the resulting vacuum energy density is 124 orders of magnitude smaller than their separate contributions, but such cancellation does not seem appealing and natural. For further details, the CCP is summarized in \cite{CC} and references therein. General considerations of CCP in the brane-world scenario can be found in \cite{Steinhardt}. In particular, a ``self-tuning" mechanism of the cosmological constant is discussed in \cite{smallCC}\cite{Kachru} under the brane-world scenario \cite{RS2}.

With regard to these two hierarchy problems, it is interesting to notice that the hierarchy $M_{Pl}/M_{EW}\sim10^{16}$ between the Planck scale and the TeV scale is roughly the \textsl{square root} of the hierarchy $M_{Pl}/M_{CC}\sim 10^{31}$ between the Planck scale and the observed cosmological constant scale. The surprising numerical coincidence prompts us to ask: Are these two hierarchy problems related? The idea that these two hierarchy problems are actually related has been pursued by one of us (PC) and other authors \cite{A,Pisin1,Pisin2}, in which some specific mechanisms were devised to induce a cosmological constant scale of 
\begin{equation}
M_{CC}=\frac{M_{EW}^2}{M_{Pl}},
\end{equation}
which is roughly the current observed scale.
More specifically, it was proposed in \cite{Pisin1,Pisin2} that the minute cosmological constant might actually be the result of a \textsl{double suppression} of the Planck scale by the \textsl{same} hierarchy factor $a=e^{-\pi kr_c}\simeq 10^{-16}$ in the RS model. That is, the expression
\begin{equation}
M_{CC}=a^2M_{Pl}
\end{equation}
might be a natural result in the RS model. Following the same spirit, in this Letter we provide an explicit mechanism that actually proves this ansatz.

Our demonstration is associated with the Casimir energy of bulk fields in the RS geometry. 
We assume that the ground value of four-dimensional effective CC, namely, the sum of its classical value as well as the divergence (or the very large term) in the calculation of vacuum energy, is identically zero provided by other mechanisms. We then demonstrate that the inevitable contribution to CC resulted from the non-trivial topology of RS geometry induced by a massive bulk fermion can be as small as the observed CC scale, $M_{CC}$, by virtue of the mass suppression. The Casimir energy of a bulk scalar field \cite{Goldberger1,Toms1,Flachi1,Saharian1,Nojiri1,Nojiri2,Knapman,Saharian2,Saharian3} or a bulk fermion field \cite{Garriga,fermion1, fermion2, Flachi2,Flachi3,Shao} in RS model have been investigated in great details. It is known that the presence of a massive bulk field would exponentially suppress the Casimir energy by a factor $e^{-2mr}$ in flat spacetime \cite{Milton}, where $r$ is the distance between two parallel plates. We will show that this is also true for the non-flat RS geometry, and the above-mentioned ansatz can be realized without fine-tuning.

It is insufficient, however, to simply demonstrate a small quantum correction to the CC in our scheme. In order to give a satisfactory analysis of the CCP in the brane-world scenario, it is necessary to consider in addition the stabilization issue in the presence of the massive bulk fermion. The modulus stabilization in the brane-world scenario is known to be intimately related to the CCP \cite{Steinhardt}. In this paper we will discuss two stabilization mechanisms under the present setup. In particular, we will show that the Goldberger-Wise mechanism can be successfully introduced in the presence of a massive bulk fermion, without spoiling the smallness of the quantum correction. 

Apart from the order of the magnitude of the cosmological constant, the \textsl{sign} is also crucial in order to give rise to an accelerating universe instead of a decelerating one. The sign of the Casimir energy has been discuss in details in \cite{Elizalde1} (see also \cite{Elizalde2}\cite{Elizalde3}), and it is pointed out that for a scalar field with periodic boundary condition, the sign is generically negative. Therefore in the present paper we consider a bulk \textsl{fermion} field with mass $m$ instead of a scalar field. As we will see, the fermionic nature of the field drastically change the sign of the Casimir energy and therefore yield a desired positive value.

There are several motivations for introducing a massive bulk fermion, summarized in \cite{Flachi3}. In particular, the bulk fermion arises naturally as the superpartner of the radion field in a supersymmetric theory, especially the string theory realization, of the brane-world scenario. In the context of particle physics, Grossman and Neubert used a massive bulk fermion to understand the neutrino mass hierarchy \cite{Grossman}. Using their result, Kitano demonstrated some bounds on the flavor changing process \cite{Kitano}. Some other phenomenological studies have invoked bulk fermions as well \cite{Chang,Dv,Georgi1,Georgi2,Lukas,Ra}.

We emphasize that we will make no attempt in explaining the traditional CCP, that is, the delicate cancellation between various vacuum energy contributions and the bare cosmological constant; rather, we focus on the possibility that an extremely small quantum correction can be produced quite naturally from a massive bulk field, provided that the infinite part (or the very large part) can be properly renormalized into the counterterms. We also do not include the graviton self-interaction, which is in fact much larger than the observed CC. Nevertheless, we consider it a step forward in demonstrating that a naturally minute quantum correction to CC is attainable without fine-tuning, in a similar spirit as in \cite{Elizalde1}.

\section{Casimir energy for a massive bulk fermion field}

The one-loop effective potential for a massive bulk fermion is \cite{Flachi3}
\begin{align}
V_{eff}=&V^R_h+V^R_va^4\notag\\
&-\frac{k^4a^4}{16\pi^2}\int_0^\infty dt~t^3~\mbox{ln}\left[1-\frac{K_\mu(t)I_\mu(at)}{I_\mu(t)K_\mu(at)}\right],\label{Veff2}
\end{align}
where $K_\mu(t)$ and $I_\mu(t)$ are the modified Bessel functions and $V_{h,v}^R$ are the shifts of the renormalized brane tension from their classical values, $V_h^0=-V_v^0=24M^3k$. The definition of $\mu$ is
\begin{equation}
\mu=\frac{1}{2}\pm \frac{m}{k},\label{mu}
\end{equation}
where the $\pm$ stands for the type I and type II boundary condition, respectively \cite{Flachi3}. (The minus sign is totally acceptable, since the Dirac spinor mass changes sign under parity.)

The small $a$ expansion of (\ref{Veff2}) reads:
\begin{equation}
V_{eff}=V^R_h+V^R_va^4+\frac{k^4a^4}{16\pi^2}f(\mu)a^{2\mu}+\mathcal{O}(a^{2\mu+2}),\label{fermion}
\end{equation}
where
\begin{equation}
f(\mu)\equiv \frac{2}{\mu\Gamma(\mu)^2}\int_0^\infty dt~t^{2\mu}\frac{K_\mu(t)}{I_\mu(t)}.
\end{equation}
Notice that $a^{2\mu}$ in (\ref{fermion}) is a significant suppression factor for generic values of $\mu=1/2\pm m/k$. This is in complete analogy to the suppression factor $e^{-2mr}$ for the Casimir energy of a massive field in flat spacetime \cite{Milton}, where $r$ is the distance between two parallel plates. 
The suppression would be effective, however, only if the function $f(\mu)$ is not exponentially large. This is indeed the case for $m/k\sim \mathcal{O}(1)$, which is a natural choice for $m$ since the 5D Planck mass $M (\sim k)$ is the only fundamental scale in the RS model. 
Some values of interest for $f(\mu)$ are listed below:
\begin{equation}
f(0.6)\sim2.1,~f(1.3)\sim8.7,~f(2.0)\sim 29.
\end{equation}
For our purpose, we choose $\mu\simeq2$, which corresponds to $m\simeq \pm 1.5k$. Then we have 
\begin{equation}
V_{eff}\simeq V^R_h+V^R_va^4+ \frac{f(2)}{16\pi^2}a^8 k^4.\label{a8k4}
\end{equation}
The last term $a^8k^4$ in (\ref{a8k4}) corresponds to a mass scale $a^2k\sim M_{CC}$, which is the desired order of magnitude for the observed CC and is obtained without fine tuning. This is the main result of this paper. It should be mentioned that as in the original RS model, the curvature $k$ is chosen to be smaller than $M$. Thus even if the fermion mass $m$ is larger than $k$, we may still choose it to be smaller than $M$.
We again emphasize that the quantum correction (the third term) in (\ref{a8k4}) is \textsl{positive} due to the fermionic nature of the field, in contrast to the negative Casimir energy for bulk scalar field found in \cite{Goldberger1}.

\section{Stabilization}
If we naively assume $V_{v,h}^R$ to be zero, i.e., the brane tensions are not shifted by a finite amount due to quantum corrections, then the value of the effective potential obtained above would indeed correspond to the observed CC. However, such potential alone cannot stabilize the orbifold radius due to its monotonic dependence on $a$. A stabilization mechanism is therefore necessary. 

\subsection{Garriga-Pujol\`as-Tanaka Mechanism}
Consider first the effective potential (\ref{a8k4}) induced solely by the massive bulk fermion \cite{Garriga}. The brane tension shifts $V^R_{v,h}$ are in general determined by the renormalization conditions:
 \begin{align}
 V_{eff}(a_{obs})=\Lambda_{obs},\notag\\
 \frac{dV_{eff}}{da}(a_{obs})=0\label{ren}.
 \end{align}
We then find, in this case,
\begin{align}
&V_h^R\simeq \frac{f(2)}{16\pi^2}a^8k^4+\Lambda_{obs},\label{Vhmu}\\
&V_v^R\simeq -\frac{f(2)}{8\pi^2}a^4k^4.
\end{align}
So the total values of the the brane tensions are
\begin{align}
V_h&=V_h^0+V_h^R\simeq (10^{19}\mbox{GeV})^4+(10^{-4}\mbox{eV})^4,\label{Vh}\\
V_v&=V_v^0+V_v^R\simeq-(10^{19}\mbox{GeV})^4-(\mbox{TeV})^4.\label{Vv}
\end{align}
It is interesting to note that the two terms in (\ref{Vhmu}) are of the same order, in contrast to that for the massless bulk fermion. However, the fine-tuning problem reappears at the stage of stabilization, as seen in (\ref{Vh}) and (\ref{Vv}). This motivates us to look for an alternative mechanism.

\subsection{Goldberger-Wise Mechanism}
In the Goldberger-Wise mechanism, a massive bulk scalar field $\Phi$ with a brane self-interaction induces a stabilizing potential \cite{GoldbergerWise}:
\begin{equation}
V_\Phi(a)=4ka^4\big(v_v-v_h a^\epsilon\big)^2,\label{Vp}
\end{equation}
where $\epsilon=m_{\Phi}^2/4k^2$ is treated as a small number and $v_{v,h}$ are of mass dimension $3/2$. This potential is added to (\ref{a8k4}) to ensure the stability.

There are two concerns about whether the introduction of the bulk scalar field $\Phi$ would overwhelm the small fermionic Casimir energy. First, the mass $m_\Phi$ of the bulk scalar field is assumed to be small compared with $k$, but still roughly of the same order. Therefore its induced Casimir energy might be larger than that by the fermion field, which is suppressed by a large mass $m\simeq 1.5k$. Second, since now the total effective potential is the sum of (\ref{a8k4}) and (\ref{Vp}), its minimum value might deviate significantly from zero. It turns out that we are actually safe from these problems.

With regard to the first concern, let us consider the Casimir energy of the bulk scalar field minimally coupled to the curvature (which is the case for \cite{GoldbergerWise}) \cite{Goldberger1},
\begin{equation}
V_{Cas,\Phi}=-\frac{k^4a^4}{16\pi^2}g(\nu)a^{2\nu}+\mathcal{O}(a^{2\nu+2}),\label{VCas}
\end{equation}
where $g(\nu)$ is some unimportant numerical constant of the order unity. We note that the Casimir energy is suppressed by the factor $a^{2\nu}$, in a similar fashion as that for the fermion. The crucial difference is that for the scalar field case,
\begin{equation}
\nu=\sqrt{4+m_\Phi^2/k^2}.\label{nu}
\end{equation}
So even for $\epsilon=m_{\Phi}^2/4k^2\ll1$ (but $m_{\Phi}\lesssim k$), $\nu$ is still slightly larger than 2, and it thus induces a Casimir energy of the \textsl{same} order as that for the fermion. That is, the Casimir energy induced by $\Phi$ does not overwhelm that by the bulk fermion. This is in contrast with the fermion case, where only a fermion with mass $\sim1.5k$ would correspond to a Casimir energy about $a^8k^4$. The origin of this difference lies in that the bulk scalar field $\Phi$ in \cite{GoldbergerWise} is \textsl{minimally} coupled to the curvature, rather than conformally coupled. Had we invoked a conformally coupled bulk scalar field, $\nu$ would have been replaced by $\sqrt{1/4+m_\Phi^2/k^2}$ \cite{Flachi1}. Then the corresponding Casimir energy would be of the order $a^5k^4$ and would dominate over the fermionic contribution.

The second concern is more straightforward. The total effective potential is
\begin{equation}
V_{eff}(a)\simeq Bk^4a^8+4ka^4\big(v_v-v_h a^\epsilon\big)^2,\label{19}
\end{equation}
where $B=\left[f(2)-g(2)\right]/(16\pi^2)$ is a numerical constant of the order unity. Note that we have combined the Casimir energies for bulk scalar and fermion fields into the first term, which will be treated as a perturbation to the second. In addition, we have set the brane tension shifts $V^R_{v,h}$ to zero. 

To the zeroth order, $a_{min}=(v_v/v_h)^{1/\epsilon}\equiv a_0\sim 10^{-16}$. Assuming that the true minimum is $a_{min}=a_0+\delta$, putting this into the derivative of (\ref{19}) and setting it to zero, we have
\begin{align}
0=&\frac{dV_{eff}}{da}(a_{min}=a_0+\delta)\notag\\
=&8Bk^4a_0^7+(56Bk^4a_0^6+8kv_h^2a_0^{2\epsilon+2}\epsilon^2)\delta+\mathcal{O}(\delta^2).
\end{align}
The first term in the parenthesis can be dropped due to its $a_0^6$ dependence. Further assuming that $v_{v,h}$ are of the Planck scale, which is the only scale for the parameters in RS model, we find $\delta\sim a_0^{5-2\epsilon}$. 

Now we put $a_{min}=a_0+\delta$ back into (\ref{19}). The minimum value of $V_{eff}$ then reads:
\begin{align}
V_{eff}=& Bk^4a_0^8+8Bk^4a_0^7\delta+\mathcal{O}(\delta^2)\notag\\
\sim &Bk^4a_0^8+Ck^4a_0^{12-2\epsilon} +\mathcal{O}(\delta^2),\label{21}
\end{align}
where $C$ is some unimportant constant of the order unity. It is clear that the second term in (\ref{21}) is much smaller than the leading term, which is indeed of the scale $M_{CC}^4$. 

\section{discussions}
In summary, we have demonstrated that a bulk fermion with mass $m\simeq \pm1.5k$ can induce a Casimir energy that is of the observed cosmological constant scale. To ensure the stability of the system, a self-interacting bulk scalar field $\Phi$ with mass $m_{\Phi}\lesssim k$ is introduced as in \cite{GoldbergerWise}. As shown above, the Casimir energy induced by $\Phi$ is of the same order as that by the fermion, and the resulting leading term of the minimum effective potential remains to be of the desired order $a^8k^4$.

It should be mentioned that by setting $V^R_{v,h}$ to zero, it actually implies that the infinite part of the brane tension shifts must be finely tuned to absorb the divergence (or very large term) in the calculation of the Casimir energies, in such a way that no residue is left behind. We made no attempt to explain this delicate cancellation in the process of renormalization, but focused only on the possibility that an extremely minute quantum correction can indeed be achieved without fine-tuning of the bulk field mass. In this regard, our philosophy is similar to that in \cite{Elizalde1}.

It would be interesting to further pursue the possible relations between the bulk fermion and scalar fields, and also their phenomenological implications. In addition, the dynamical origin for such massive bulk fermion field is relevant and should be further investigated.

\begin{acknowledgments}
We thank D. Maity and A. Flachi for their useful suggestions and comments on the subject. We are also grateful to K. Y. Su and C. I. Chiang for interesting and encouraging discussions. This research is supported by Taiwan National Science Council under Project No. NSC 97-2112-M-002-026-MY3, by Taiwan's National Center for Theoretical Sciences (NCTS), and by US Department of Energy under Contract No. DE-AC03-76SF00515.
\end{acknowledgments}

\end{document}